\documentclass[10pt,a4paper]{article}

\usepackage[T1]{fontenc}
\usepackage[utf8]{inputenc}
\usepackage{lmodern}
\usepackage{microtype}
\usepackage[margin=2.4cm]{geometry}
\usepackage{amsmath,amssymb,amsthm}
\usepackage{graphicx}
\usepackage{booktabs}
\usepackage{setspace}
\usepackage[labelfont=bf,font=footnotesize]{caption}
\usepackage[numbers,sort&compress]{natbib}
\usepackage{xcolor}
\usepackage{xurl}
\usepackage[colorlinks=true,linkcolor=black,citecolor=blue!50!black,
            urlcolor=blue!50!black]{hyperref}

\usepackage{mathtools}
\usepackage{bm}
\usepackage{multirow}
\usepackage{array}
\usepackage{enumitem}
\usepackage{algorithm}
\usepackage{algpseudocode}
\graphicspath{{figures/}}

\newcommand{\PaperTitle}{Accurate structural identification for layered composite structures, through a wave and finite element scheme}
\newcommand{\PaperAuthors}{D. Chronopoulos$^{a}$, C. Droz$^{b}$, R. Apalowo$^{a}$, M. Ichchou$^{c}$ and W. J. Yan$^{a,d}$}
\newcommand{\PaperAffil}{%
  $^{a}$Institute for Aerospace Technology \& The Composites Group, The University of Nottingham, NG7 2RD, UK\\
$^{b}$Katholieke Universiteit Leuven, Noise and Vibration Research Group, Division PMA - Celestijnenlaan 300 Leuven, Belgium\\
$^{c}$Ecole Centrale de Lyon, 36 Avenue Guy de Collongue, 69134 Ecully Cedex, France\\
$^{d}$Department of Civil and Hydraulic Engineering, Hefei University of Technology, Hefei, Anhui 23009, PR China}
\newcommand{\PaperDate}{Accepted manuscript, originally 2017}
\newcommand{\PaperKeywords}{Structural identification, Non-Destructive Evaluation, Finite Elements, Wave Propagation, Layered Structures, Ultrasound}

\newcommand{\PaperDOI}{10.1016/j.compstruct.2017.09.062}
\newcommand{\HalID}{hal-03372998}
\newcommand{\ArxivID}{}          

\newcommand{\CiteAs}{%
  D. Chronopoulos, C. Droz, R. Apalowo, M. Ichchou, W. J. Yan, \emph{\PaperTitle}, Composite Structures 182 (2017) 566-578, doi:10.1016/j.compstruct.2017.09.062}

\makeatletter
\newcommand{\HasDOI}{%
  \ifx\PaperDOI\@empty\expandafter\@secondoftwo\else\expandafter\@firstoftwo\fi}
\newcommand{\HasArxiv}{%
  \ifx\ArxivID\@empty\expandafter\@secondoftwo\else\expandafter\@firstoftwo\fi}
\makeatother

\providecommand{\doi}[1]{\href{https://doi.org/#1}{doi:#1}} 

\begin{document}

\begingroup
\renewcommand{\thefootnote}{\fnsymbol{footnote}}
\begin{center}
{\LARGE\bfseries \PaperTitle\par}
\vspace{0.7em}
{\normalsize \PaperAuthors\par}
\end{center}
\footnotetext[1]{\raggedright\PaperAffil}
\footnotetext[2]{\PaperDate.
  Author's version with AI-assisted typesetting and layout. Not the publisher version of record.
  \HasDOI{Definitive publication: \doi{\PaperDOI}.}{No journal DOI yet.}
  HAL: \href{https://hal.science/\HalID}{\HalID}.
  \HasArxiv{arXiv: \href{https://arxiv.org/abs/\ArxivID}{\ArxivID}.}{}}
\endgroup
\setcounter{footnote}{0}

\vspace{0.85em}
{\small
\noindent\textit{How to cite this version.}\\[0.35em]
\raggedright \CiteAs
\par}
\vspace{1.05em}

\begin{abstract}
\noindent
We present for the first time an approach for identifying the geometric and material characteristics of layered composite structures through an inverse wave and finite element approach. More specifically, this Non-Destructive Evaluation (NDE) approach is able to recover the thickness, density, as well as all independent mechanical characteristics such as the tensile and shear moduli for each layer of the composite structure under investigation. This is achieved through multi-frequency single shot measurements. It is emphasized that the success of the approach is independent of the employed excitation frequency regime, meaning that both structural dynamics and ultrasound frequency spectra can be employed. It is demonstrated that more efficient convergence of the identification process is attained closer to the bending-to-shear transition range of the layered structure. Since a full FE description is employed for the periodic composite, the presented approach is able to account for structures of arbitrary complexity. The procedure is applied to a sandwich panel with composite facesheets and results are compared with two wave-based characterization techniques: the Inhomogeneous Wave Correlation method and the Transition Frequency Characterization method. Numerical simulations and experimental results are presented to verify the robustness of the proposed method.
\end{abstract}



\section*{Nomenclature}\label{sec:nomenclature}

\noindent
\begin{tabular}{@{}p{0.2\linewidth}p{0.76\linewidth}@{}}
$\mathbf{M}$, $\mathbf{K}$ & mass and stiffness matrices of the periodic waveguide \\
$\mathbf{D}$ & dynamic stiffness matrix (DMS) of a waveguide's modelled periodic segment \\
$\mathcal{F}$ & objective function to be minimized \\
$\mathbf{T}$ & transfer matrix of the wave propagation eigenproblem \\
$\mathbf{f}$ & forcing vector for an elastic waveguide \\
$\mathbf{p}$ & vector of structural characteristics to be identified \\
$\mathbf{q}$ & physical displacement vector for an elastic waveguide \\
$\mathbf{x}(t)$ & logged signal vector as a function of time \\
$\rho$, $h$, $E$, $G$, $v$ & mass density, thickness and mechanical characteristics of each layer \\
$L_x$ & dimension of a waveguide's modelled periodic segment \\
$L$, $R$, $I$ & left, right sides and interior indices \\
$U_0$ & amplitude of applied excitation signal \\
$c_p$ & wave phase velocity \\
$c_g$ & wave group velocity \\
$f_0$ & frequency of the applied excitation \\
$k$ & wavenumber \\
$l$, $l_{max}$ & index corresponding to layer number and total number of structural layers \\
$m$, $m_{max}$ & index corresponding to each measured frequency and total number of measured frequencies \\
$n_0$ & number of cycles for the Hanning windowed excitation \\
$rf$, $fe$ & indices denoting wave characteristics obtained through measurements and the WFE scheme respectively \\
$s$ & periodic segment positioning index \\
$t$ & time \\
$x_0$, $x_1$ & coordinates of the excitation and monitoring locations on the waveguide \\
$\Phi_q^{\omega,+}$, $\Phi_q^{\omega,-}$ & grouped displacement eigenvectors for the positive and negative going elastic waves at frequency $\omega$ \\
$\Phi_f^{\omega,+}$, $\Phi_f^{\omega,-}$ & grouped forcing eigenvectors for the positive and negative going elastic waves at frequency $\omega$ \\
$\boldsymbol{\phi}_q$, $\boldsymbol{\phi}_f$ & displacement and forcing eigenvectors \\
$\beta$ & arbitrary structural property \\
$\gamma$ & propagation constant and eigenvalue of the wave propagation eigenproblem \\
$\omega$ & angular frequency \\
\end{tabular}

\section{Introduction}\label{sec:1}

Composites are widely used in modern industry, due to their low density and high dynamic and static performances. This goal has led to develop new sandwich structures and composite materials in general, with tailored properties and a wide range of possible configurations and topologies. However, the verification and Non-Destructive Evaluation (NDE) of the actual mechanical properties of the assembled layered structure remains a very much open engineering challenge. Experimental testing and system identification have played important roles in various fields such as civil engineering, mechanical engineering and aerospace engineering due to their versatile applications such as assessing system conditions and reconciling numerical predictions with experimental investigations \cite{ref1,ref2,ref3,ref4,ref5,ref6}. In a broad context, 'system identification' refers to the extraction of information about the system behavior directly from experimental data \cite{ref7,ref8}. Over the past decades, different system identification methods in the time domain \cite{ref9,ref10,ref11}, frequency domain \cite{ref12,ref13} and time-frequency domain \cite{ref14,ref15} have been proposed. System identification has been applied extensively in the field of structural dynamics and it has been proven to be useful in the analysis of the dynamic behavior of the structure. In the context of structural dynamics, system identification generally includes modal-parameter identification by extracting the modal data of a structural system such as its natural frequencies, damping ratio and mode shapes as well as physical-parameter identification by extracting useful information related to stiffness, mass and damping. Numerous approaches have been developed for system identification including stochastic subspace identification method \cite{ref16}, extended Kalman filter method \cite{ref17} and Bayesian approaches \cite{ref18,ref19,ref20} to cite a few of them.

The system identification approaches aforementioned are generally based on the measurement of structural vibration information. Nowadays however, several researchers have shown that propagating wave properties can have a higher sensitivity to structural parameters than other structural responses. Therefore, sporadic but consistent efforts have been directed to extract a system's structural condition using wave propagation information over the past decade \cite{ref21,ref22}. However, it is worth mentioning here that rare works reviewed in \cite{ref21,ref22} are dependent on the model. Though some efforts \cite{ref23,ref24,ref25} have been devoted to inferring the model parameters through wave propagation, they have not resulted in full-fledged applications. Therefore, there is still significant room for further exploration in system identification by integrating mathematical models of wave propagation. Many methods have been developed to perform material characterization in composites. One can cite the experimental method for the characterization of Nomex cores \cite{ref26}, or the vibratory identification technique proposed in Matter et al. \cite{ref27}. Other methods based on numerical strategies were also developed in \cite{ref28,ref29}. Recently, a Transition Frequency Characterization technique \cite{ref30} was developed to perform material identification in sandwich structures, based on the so-called bending-to-shear conversion effect. However, such methods could not handle more complex topologies often encountered in transportation industry, despite the considerable progress made on the numerical wave-based models in this field.

The propagation of guided waves in sandwich structures has indeed been the subject of intense research in the recent years. Traditional analytical methods (i.e. classical plate theory, Mindlin type or first-order shear deformation theories) typically employed for modelling wave propagation in monolayers can only correctly capture the wave characteristics in the low frequency range for thick structures. In contrast, Finite Element (FE) based wave methods assume a full 3D displacement field and are therefore capable of capturing the entirety of wave motion types in the waveguide under investigation in a very accurate and efficient manner. FE-based wave propagation within periodic structures was firstly considered in the pioneering work of the author of \cite{ref31}. The work was extended to two dimensional media in \cite{ref32}. The Wave and Finite Element (WFE) method was introduced in \cite{ref33,ref34} in order to facilitate the post-processing of the eigenproblem solutions and further improve the computational efficiency of the method, while the extension of WFE method for two dimensional structures was introduced in \cite{ref35}.

The principal novel contribution of this work is the development of a comprehensive methodology coupling periodic structure theory to FE in order to identify the characteristics of each individual layer of a composite structure through experimental measurements on the entire structure. The method is robust and can account for structures of arbitrary complexity. Both low as well as high frequency excitations can be employed for inverting the structural problem. It is shown that faster convergence can be acquired around the wave transition region \cite{ref36,ref37} which is a specific type of wave conversion \cite{ref38,ref39}, occurring in sandwich structures subject to flexural vibrations. Both experimental, as well as numerical case studies are presented in order to validate the exhibited methodology.

The paper is organised as follows: In Section 2 the FE computational scheme for predicting wave propagation in multilayered structures is presented and targeted suggestions are made in order to effectively recover the structural and material characteristics for the structure under investigation. A Hilbert Transform is employed to measure the time of arrival of the wave pulses and subsequently the propagating wavenumbers. A Newton-like iterative scheme is eventually employed for minimising the formulated objective function and recovering the mechanical characteristics of each individual layer through solution of the system of eigenvalue expressions. In Section 3 several experimental and numerical case studies are presented for validating the exhibited identification approach. A periodic layered structure is modelled and multi-frequency single wave shots are excited and measured. The structural and material characteristics for each layer are then recovered. Conclusions are eventually drawn in Section 4.

\section{An inverse wave and finite element methodology for structural identification}\label{sec:2}

Mathematical modeling can provide a good understanding and form the basis of a characterization process for a mechanical system. Given the mathematical model, system identification can be implemented by fitting it to that from experimental testing. In the present paper, the primary focus is to improve structural models by measurements performed on the real structure using wave propagation measurement data. As a result, one can make inference about the parameters of a mathematical model based on the observed measurements.

\begin{figure}[htbp]
  \centering
  \includegraphics[width=0.4\columnwidth]{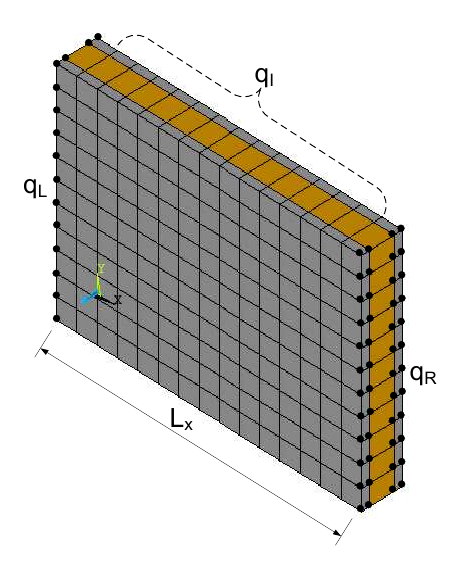}
  \caption{Caption of the WFE modelled composite waveguide with the left and right side nodes $\mathbf{q}_L$, $\mathbf{q}_R$ bullet marked. The range of interior nodes $\mathbf{q}_I$ is also illustrated.}
  \label{fig:1}
\end{figure}

An arbitrarily complex and periodic in the $x$ direction waveguide is illustrated in Fig.~\ref{fig:1}. The structure may comprise an arbitrary number of layers which may be anisotropic. It is assumed that some of the structural characteristics are unknown (or even variable over time) and need to be evaluated through a non-destructive evaluation process. The identifiable properties include the thickness, density as well as the material characteristics of each individual layers. In the following, a wave and finite element scheme is employed in order to recover the required properties of the layered structure through the acquired propagating wave data.

\subsection{Obtaining the reference wave characteristics}\label{sec:2.1}

The required data to be extracted and later fed into the structural identification process are the wave phase speeds (or wavenumbers) of specific wave types propagating within the laminate under investigation. A number of methods can be employed for exciting and measuring specific propagating wave modes within a composite structure. Piezoelectric \cite{ref40} or even non-contact laser actuation \cite{ref41} can be employed for exciting and measuring wave properties in the ultrasound frequency range, while within the structural dynamics spectrum more conventional shaker and accelerometer devices can be employed (see also the experimental study presented in Section 3).

With regard to numerical calculations, a number of approaches can be employed \cite{ref42} for exciting the structure and computing its response at any node. Different wave types can be excited by employing their corresponding displacement field. Care has to be taken in order to ensure that a sufficiently fine discretization (at least 6-10 elements per wavelength) has been employed for correctly capturing the propagating waves in the layered structure.

It is hereby assumed that the excitation frequency of the wave packages is known and can be controlled as well as altered within a certain spectrum. It is generally beneficial to have a range of well separated excitation frequencies (by at least 50\% from each other) in order for the post-processing identification process to converge at a faster rate. To avoid frequency leakage, a proper signal windowing technique should also be employed. A Hanning window was chosen as the most appropriate and was employed throughout the results presented in this manuscript. The excitation signals are quasi-monochromatic burst of amplitude $U_0$, centred around frequency $f_0$ and involving a number of $n_0$ cycles. Input signals are windowed so that the input signal is defined by $u(t) = U_0 \sin\left(\frac{\pi f_0 t}{n_0}\right)\sin(2\pi f_0 t)$ for $0 \le t \le \frac{n_0}{f_0}$ and $u(t) = 0$ for $t > \frac{n_0}{f_0}$.

\begin{figure}[htbp]
  \centering
  \includegraphics[width=\linewidth]{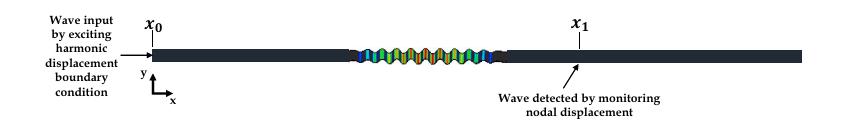}
  \caption{Illustration of the suggested configuration for obtaining the reference wave characteristics to be later compared with the WFE ones. All simulations are performed using ANSYS V4.15. Three-dimensional solid brick elements are employed for enhanced accuracy and a minimum mesh density of 15 elements per wavelength is retained.}
  \label{fig:2}
\end{figure}

An illustration of the configuration is depicted in Fig.~\ref{fig:2}. The waveguide is excited at a specified central harmonic signal of frequency $f_0$ at a location $x = x_0$ and the signal is monitored at location $x = x_1$, after which the signal has travelled over a distance of $L = x_1 - x_0$. Once the experimental or numerical signal measurements are logged, the wavenumbers and group velocities of the excited waves can be easily determined.

Time histories are initially registered at the excitation and monitoring locations. The maximum amplitudes of the time history signals $\mathbf{x}(t)$ are obtained from the Hilbert Transforms $H[\mathbf{x}(t)]$ of the acquired signals in the time domain. Hilbert Transform $H[\mathbf{x}(t)]$ of the acquired time signal $\mathbf{x}(t)$ is used to evaluate the main attributes of $\mathbf{x}(t)$. The signal envelope is determined at emission, $X_0$ and arrival, $X_1$ while the time delay is defined by the time difference between the maximal amplitudes of the envelopes. The total Time of Flight of the wave signal from the point of excitation to the monitoring point is measured as the time difference $t(x_1) - t(x_0)$ between the maximum amplitudes of the excited and the monitored signal envelopes. In ultrasonic NDE, the wavenumber of the wave package is straightforward to obtain as:
\begin{equation*}
k = \frac{\omega}{c_{p}}
\tag{1}\label{eq:1}
\end{equation*}
where the phase velocity of the signal $c_p$ can be obtained from its ToF and its propagation distance $L$. It is noted that the phase velocity for a non-dispersive wave is equal to its group velocity.

\subsubsection{Optimal excitation frequency range for the reference characteristics}\label{sec:2.1.1}

As aforementioned, the exhibited scheme can be employed for both in the high as well as the lower frequency range. It is interesting to note that sandwich laminates comprising a soft core (e.g. honeycomb or foam one), generally exhibit a so-called transition frequency with regard to their propagating flexural wave speed behaviour.

\begin{figure}[htbp]
  \centering
  \includegraphics[width=0.80\linewidth]{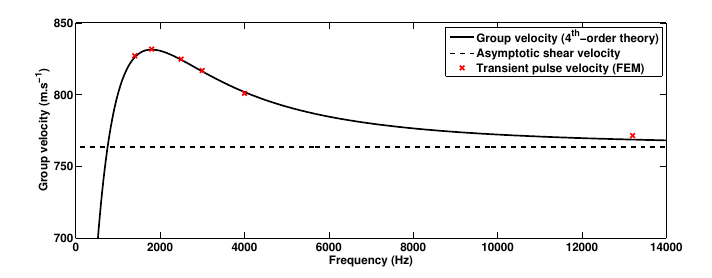}
  \caption{Comparison between theoretical dispersion curves and the velocities obtained by ToF with transient pulse simulations at various frequencies. The transition frequency of the sandwich panel is clearly distinguished and characterized by a maximum value for the group velocity. (from \cite{ref30})}
  \label{fig:3}
\end{figure}

Kurtze and Watters \cite{ref43} were the first to observe and develop an asymptotic model for the wave dispersion into symmetric flat thick sandwich structures. They divided the flexural wave speed of a sandwich panel (frequency-wise) into three sections, the first characterized by the panel vibrating as a whole, the second by the core's shear wave speed and the third by each of the two facesheets vibrating separately and loaded with half of the core mass. In the same work, it was also shown that the flexural wave type in the sandwich structure has its maximum group velocity value at the transition frequency. As an illustrative example, the group velocities for a typical sandwich structure (to be investigated in Section 3) are computed and compared with the transient simulations in Fig.~\ref{fig:3} (see \cite{ref30} for details of the conducted simulation). Since the transition frequency of a sandwich composite is intensely sensitive to its structural characteristics, it becomes evident that exciting the structure under investigation close to this frequency range can unveil its intrinsic properties and facilitate the convergence of the algorithm.

It is noted that this transition for most sandwich laminated occurs well within their 'structural dynamics' response range (typically between 2 kHz and 10 kHz). If higher frequencies are excited for the identification process (e.g. ultrasound range) then obviously exciting the transition range is not feasible, however the identification process will still converge to the correct properties as shown in Section 3.

\subsection{Structural identification methodology}\label{sec:2.2}

\subsubsection{Considerations on the forward wave FE model}\label{sec:2.2.1}

Once again, we consider the periodic complex waveguide illustrated in Fig.~\ref{fig:1}. The propagation constants for the elastic waves travelling in the x direction can be sought through the forward Wave and Finite Element (WFE) scheme as described in Appendix A. It is noted that analytical multilayer modelling techniques \cite{ref44} have also shown to successfully predict the broadband wave properties for layered structures; however a FE based method is hereby preferred thanks to its versatility (different numbers of layers and complex material properties are straightforward to take into account, with no need of altering the modelling approach). Moreover, 3D displacement fields are employed, therefore retaining accuracy in a broadband frequency regime without any implicit strain field assumptions (such as the ones used by shear deformation models \cite{ref45}).

It is obvious that the number of solutions for the formulated eigenvalue problem depends on its size. Most of the obtained solutions however correspond to either fast decaying, evanescent waves or to numerical artefacts which bear no physical value. By employing straightforward filtering approaches (typically by comparing the real and imaginary parts of the obtained wavenumbers as in \cite{ref46}) the propagating and positive-going wave types can be distinguished and kept in a separate database for later comparing them with the acquired reference wave characteristics. Using the obtained wave mode shapes $\boldsymbol{\phi}_q$ (see also Appendix A) the actual wave type can also be categorized (typically flexural, shear or longitudinal) in order to ensure the fact that the user is always comparing WFE and reference values for the same wave type.

Due to the large number of numerical artefacts (especially for large models), it is challenging if not impossible to experimentally reconstruct the vector of eigenvalues and attempt a direct inversion \cite{ref47} of the eigenvalue problem of Appendix A (Eq. (A.8)). Moreover, given the experimental capability of individually exciting just one or two propagating wave types in a structure it is oftentimes only possible to recover a single eigenvalue per frequency for the eigenproblem of Eq. (A.8). On the other hand, the employment of wave based measurements suggests that eigenvalue data for an unlimited number of frequencies can be extracted, which is the principal advantage of the suggested method compared to modal based identification techniques.

\subsubsection{Formulation of the identification objective function}\label{sec:2.2.2}

The advantage of the exhibited WFE approach is therefore the fact that since the excitation frequency is controlled and known, an unlimited number of eigenvalues (for the same wave) can be extracted for the corresponding number of frequencies. Since each resulting $j_{\text{th}}$ eigenvalue (propagation constants for each wave type) can be expressed as
\begin{equation*}
\gamma_{j,fe} = \mathrm{e}^{-\mathrm{i} k_{j,fe} L_{x}}
\tag{2}\label{eq:2}
\end{equation*}
the corresponding wavenumber $k_j$ can be given by
\begin{equation*}
k_{j,fe} = \frac{\log \gamma_{j,fe}}{-\mathrm{i} L_{x}}
\tag{3}\label{eq:3}
\end{equation*}
which can be directly compared to the reference wavenumber values $k_{j,rf}$. The objective function of the identification process to be minimized is then obtained through a least squares approach as
\begin{equation*}
\mathcal{F}(\mathbf{p}) = \sum_{m=1}^{m_{max}} \left( k_{m,rf} - k_{m,fe} \right)^{2}
\tag{4}\label{eq:4}
\end{equation*}
with $k_{m,rf}$ and $k_{m,fe}$ being measured and calculated respectively at frequency $\omega_m$ for the same wave type, while $\mathbf{p}$ is the vector of parameters to be identified; in the very general case this is expressed as
\begin{equation*}
\mathbf{p} = \left\{ E_{x,1} E_{y,1} E_{z,1} v_{xy,1} v_{xz,1} v_{yz,1} G_{xy,1} G_{xz,1} G_{yz,1} h_1 \rho_{m,1} \cdots \rho_{m,l_{max}} \right\}^{\top}
\tag{5}\label{eq:5}
\end{equation*}
for layers $l \in [1, l_{max}]$. In the above, $m_{max}$ is the total number of reference eigenvalues which can be used in the identification procedure. It is obvious that the minimum required $m_{max}$ is equal to the number of parameters to be identified, however results for additional frequencies will generally improve the precision of the identification process. An excessive $m_{max}$ is undesired, as for each computation of $\mathcal{F}$, an equivalent number of eigenproblems needs to be solved.

In order to accelerate the Newton-like iterative scheme, the first (or even the second) gradient of the objective function $\frac{\partial \mathcal{F}}{\partial \beta_i}$ may be provided for each sought structural property $\beta_i$ as
\begin{equation*}
\frac{\partial \mathcal{F}}{\partial \beta_{i}} = \sum_{m=1}^{m_{max}} \left( 2 k_{m,fe} \frac{\partial k_{m,fe}}{\partial \beta_{i}} - 2 k_{m,rf} \frac{\partial k_{m,fe}}{\partial \beta_{i}} \right)
\tag{6}\label{eq:6}
\end{equation*}

It is noted that the set of parameters may be considered to have constrained values (e.g. $\beta_i \in [\beta_{i,min}, \beta_{i,max}]$), again for practical reasons. In order to compute the gradient of the wavenumber values $\frac{\partial k_{m,fe}}{\partial \beta_i}$, an expression of the mass and stiffness matrices directly as a function of the material and geometric characteristics of each layer is greatly practical, as derived in Appendix B. By employing the symbolic expressions for the mass and stiffness derivatives $\frac{\partial \mathbf{M}}{\partial \beta_i}$, $\frac{\partial \mathbf{K}}{\partial \beta_i}$ the wavenumber sensitivity $\frac{\partial k_{m,fe}}{\partial \beta_i}$ can be computed as in \cite{ref48}.

The constrained minimization problem can be implemented within standard mathematics software and nonlinear optimization algorithms (such as fmincon in MATLAB) can be employed in order to compute the optimal parameter vector $\mathbf{p}$ that minimizes $\mathcal{F}(\mathbf{p})$ and corresponds to the identified structural properties. It is stressed that due to the existence of several local minima in $\mathcal{F}$, a global search algorithm should be employed during the solution process. The minimum of the encountered solutions is retained as the global set of acquired structural characteristics.

The presented scheme is validated in the following Section through numerical, as well as experimental results. It is shown that when clear wavenumber measurements are obtained, the approach can be exceptionally accurate. Moreover, the procedure can be applied within a rational amount of time (especially if only one or two structural parameters are to be sought) using conventional low-cost computing equipment. The generic iterative procedure of the post-processing identification process is presented in Algorithm 1.

\begin{algorithm}[htbp]
\caption{Newton-like iterative scheme for identifying the parameters of a layered structure}
\label{alg:1}
\begin{algorithmic}[1]
\State Input measured reference wave characteristics. Determine total number of local minima to be investigated and evaluated. Define identification criterion for objective function $\mathcal{F}$
\State $i \leftarrow 1$ Input structural parameters for initial design to be evaluated
\State Substitute new set of structural parameters in symbolic expressions of $\mathbf{M}$, $\mathbf{K}$, $\frac{\partial \mathbf{K}}{\partial \beta_i}$, $\frac{\partial \mathbf{M}}{\partial \beta_i}$ and formulate the corresponding matrices for the periodic unit cell of the layered design under investigation
\State Solve the eigenproblem of Eq. (A.8) for design $i$. Compute WFE wave velocities and wavenumbers
\State Compute $\mathcal{F}$ and the sensitivity values $\frac{\partial \mathcal{F}}{\partial \beta_i}$ for each structural parameter $\beta_i$ to be recovered
\If{$\mathrm{d}\mathcal{F} <$ Solution convergence criterion}
\State Solution corresponds to a local minimum
\If{$\mathcal{F} <$ Identification criterion}
\State Solution corresponds to global identification solution and process can end
\Else
\State Radically alter the structural parameters and go to Step 3
\EndIf
\Else
\State Use $\frac{\partial \mathcal{F}}{\partial \beta_i}$ in order to alter structural parameters for converging towards a local minimum. $i \leftarrow i+1$ (next solution step). Go to Step 3
\EndIf
\end{algorithmic}
\end{algorithm}

\begin{figure}[htbp]
  \centering
  \includegraphics[width=0.85\linewidth]{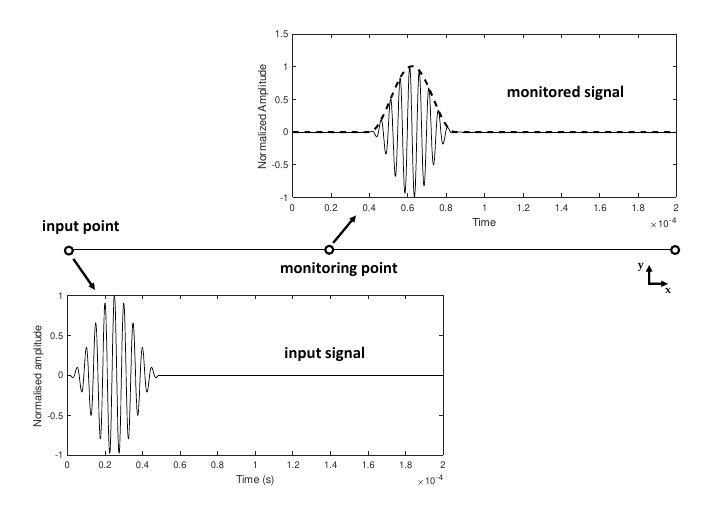}
  \caption{General representation of the ToF measurements. The pulse input is generated using an excitation device at the input point while the time delay is measured at the monitoring point. Note that better results are obtained when no edge reflections are interfering with the registered pulse.}
  \label{fig:4}
\end{figure}

\section{Numerical and experimental case studies}\label{sec:3}

\subsection{Numerical validation of the identification scheme}\label{sec:3.1}

\subsubsection{Monolayer case study}\label{sec:3.1.1}

The first numerical case study relates to identifying the thickness, density and Young's modulus for a monolayer metallic structure under investigation. The properties exhibited under Structure I (see Table~\ref{tab:1}) are employed for the monolayer case-study which is modelled by 3D solid elements in ANSYS V14.5. A longitudinal pressure wave excitation is numerically imposed at a cross section of the modelled structure. A general presentation of the measurement process is depicted in Fig.~\ref{fig:4}).

\begin{figure}[htbp]
  \centering
  \includegraphics[width=0.85\linewidth]{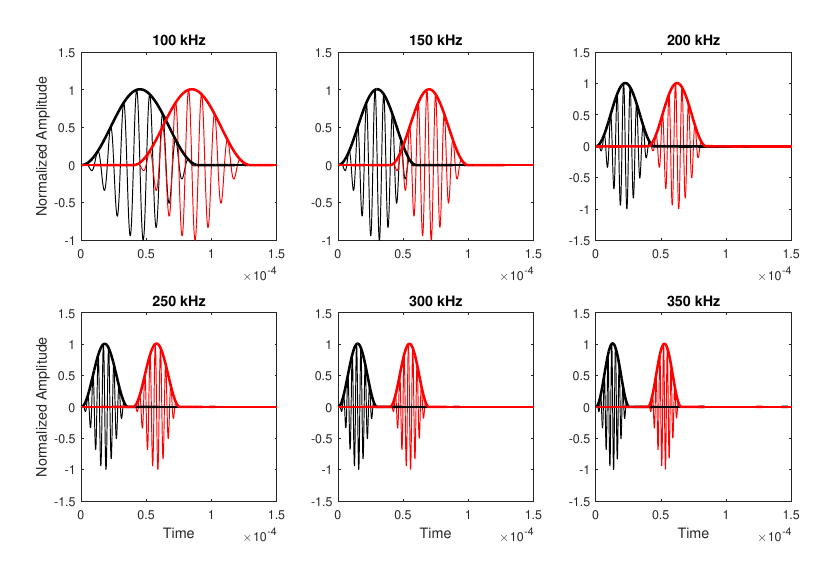}
  \caption{Time acquisition at $x = 0$ (black curves) and $x = 3$ cm (red curves) with the wave envelopes depicted in the monolayer structure. The number of cycles is $n_0 = 9$. The ToF is measured at the maximal amplitude of Hilbert transform (solid lines) signal.}
  \label{fig:5}
\end{figure}

The propagating waveform is depicted in Fig.~\ref{fig:5} for six wave pulses of different frequencies. As expected, negligible dispersion occurs for all six pulses, thanks to the high number of cycles $n_0$ employed for the Hanning window process, as well as to the non-dispersive nature of pressure waves. It is clear that the absence of reflected and converted waves at high frequencies allows a reliable determination of the wave envelope characteristics. It is stated that a quasimonochromatic burst is hereby assumed as excitation which may however not always be realistic. For structures under real operating conditions a number of impediments may exist. These could be related to the quality and feasible amplitude of the signal (good signal to noise ratio is needed) as well as to the excitation bandwidth which in reality is never monochromatic. Moreover, operational conditions (mainly temperature as well as pressurization effects) may alter the wave propagation properties of a certain modelled structure. As is the case with several fields of applied research, a comprehensive uncertainty quantification needs to be performed before applying this methodology under operational conditions, in order to determine the degree of certainty and confidence for the extracted results.

\begin{figure}[htbp]
  \centering
  \includegraphics[width=0.60\linewidth]{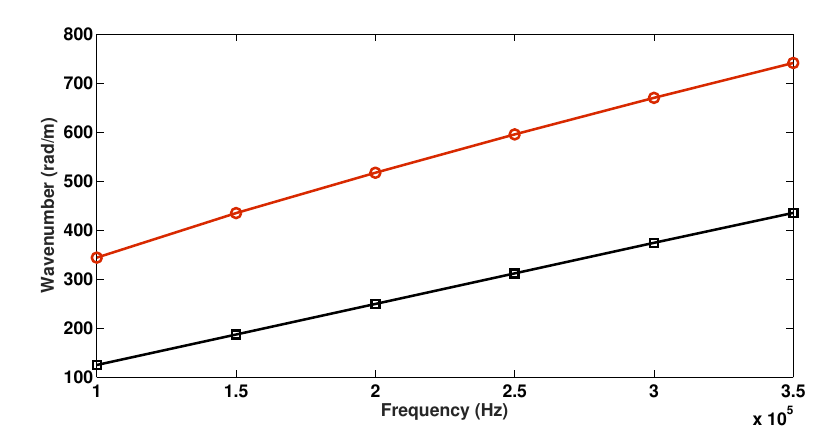}
  \caption{Reference wavenumber values obtained through a numerical solution of the full FE model for the monolayer structure. Results for a pressure propagating wave ($\square$) and a flexural propagating one ($\bigcirc$).}
  \label{fig:6}
\end{figure}

Six wavenumber measurements are recovered for an equivalent number of different ultrasonic frequencies, namely from 100 kHz to 350 kHz with a step of 50 kHz. The reference wave characteristics related to the recovered wavenumber values are shown in Fig.~\ref{fig:6}. These are retained for comparison with the WFE obtained results for the same propagating wave type which will form the objective function of the identification problem. The same process is repeated for a flexural wave propagating within the monolayer structure with the results also presented in the same Figure.

Once the reference wave characteristics $k_{m,rf}$ are established, the objective function $\mathcal{F}$ can be established as a function of the structural properties to be identified $E$, $\rho$ and $h$. A single element is employed for the formulation of the WFE model which results in very fast eigenproblem solutions for Eq. (A.8). An identification criterion equal to 10 is employed (suggesting that any local minimum with a value less than that would be considered as a solution). The minimization process was completed in 58 iterations each of which lasted approximately 8 s, resulting in a total computation time of 460s on a conventional laptop device. This suggests that employing dedicated optimization software and high-performance computing equipment would radically reduce this amount of post-processing. The final value of the objective function when pressure wave measurements were employed was of the order of $10^{-1}$. The second best identified solution gave an objective function value at the order of $10^{1}$, therefore confirming the optimality of the result. The identified parameters are exhibited in Table~\ref{tab:1} and are in excellent agreement with the ones initially used in the full FE model (maximum divergence is considerably less than 1\%). The result therefore validates the accuracy and robustness of the proposed scheme. It is noted that in theory, the Poisson's ratio of the structure could also be identified; however due to the low sensitivity exhibited by the propagating wavenumbers to this structural parameter, the time required for the process to successfully converge is much greater.

\begin{table}[htbp]
\centering
\caption{Properties of numerically modelled structural layers and identified characteristics through the inverse WFE scheme.}
\label{tab:1}
\begin{tabular}{llll}
\toprule
Structure I & Structure II & Structure III & Structure IV \\
\midrule
$\rho = 7850$ kg/m$^3$ & $\rho = 3500$ kg/m$^3$ & $\rho = 150$ kg/m$^3$ & $\rho = 3500$ kg/m$^3$ \\
$h = 1$ mm & $h = 0.5$ mm & $h = 5$ mm & $h = 1$ mm \\
$E = 170$ GPa & $E_x = 150$ GPa & $E_x = 95$ MPa & $E_x = 150$ GPa \\
-- & $E_y = 85$ GPa & $E_y = 95$ MPa & $E_y = 85$ GPa \\
-- & $E_z = 85$ GPa & $E_z = 150$ MPa & $E_z = 85$ GPa \\
$v = 0.29$ & $v_{xy} = 0.15$ & $v_{xy} = 0.3$ & $v_{xy} = 0.15$ \\
-- & $v_{xz} = 0.1$ & $v_{xz} = 0.23$ & $v_{xz} = 0.1$ \\
-- & $v_{yz} = 0.1$ & $v_{yz} = 0.3$ & $v_{yz} = 0.1$ \\
-- & $G_{xy} = 15$ GPa & $G_{xy} = 20$ MPa & $G_{xy} = 15$ GPa \\
-- & $G_{yz} = 23$ GPa & $G_{yz} = 55$ MPa & $G_{yz} = 23$ GPa \\
-- & $G_{xz} = 15$ GPa & $G_{xz} = 35$ MPa & $G_{xz} = 15$ GPa \\
\multicolumn{4}{c}{Identified structural characteristics of each layer} \\
$\rho = 7857.43$ kg/m$^3$ & $\rho = 3474.8$ kg/m$^3$ & $G_{xz} = 35.44$ MPa & $h = 1.0038$ mm \\
$h = 0.9973$ mm & -- & -- & $E_x = 148.91$ GPa \\
$E = 174.32$ GPa & -- & -- & -- \\
\bottomrule
\end{tabular}
\end{table}

\subsubsection{Layered composite case study}\label{sec:3.1.2}

\begin{figure}[htbp]
  \centering
  \includegraphics[width=0.60\textwidth]{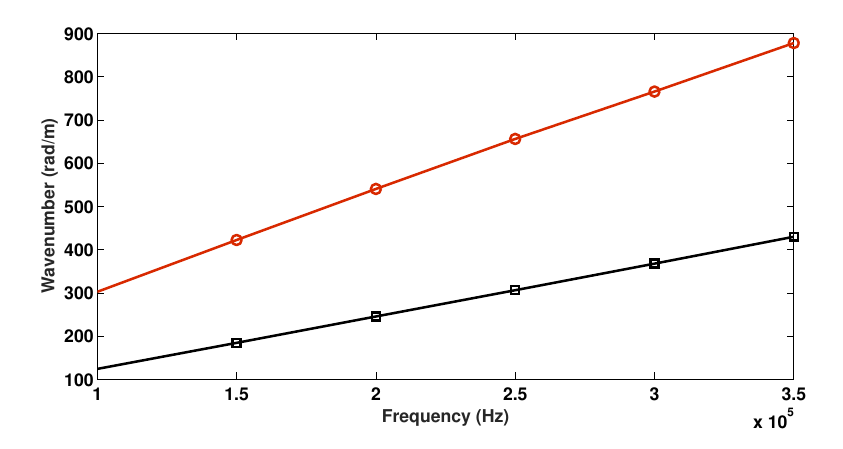}
  \caption{Reference wavenumber values obtained through a numerical solution of the full FE model for the multilayer structure. Results for a pressure propagating wave ($\square$) and a flexural propagating one ($\bigcirc$).}
  \label{fig:7}
\end{figure}

A similar process was followed for extending the calculations to a layered composite structural configuration. The properties for each layer are given as Structure II (lower facesheet), Structure III (core layer) and Structure IV (upper facesheet) in Table~\ref{tab:1}. Once again, a pressure as well as a flexural wave excitation was imposed at a specific cross-section of the modelled structure. A Hanning window was applied at all pulses with $n_0 = 9$. The results for six wave pulses of different frequencies are presented in Fig.~\ref{fig:7}.

In order to find out the maximum number of parameters that can be identified within a rational amount of time for the multilayer structure, we run the identification procedure for three, four, five and six unknown parameters. The WFE model this time comprised three FEs (one for representing each structural layer). The identification criterion was again set equal to 10. It was observed that the processes with five and six unknown parameters never converged to a satisfactory value of the objective function after 18,000 s of post-processing time (5 h). This is due to the existence of an important number of local minima that needed to be investigated by the fmincon algorithm. None of the derived local solutions however had a value close to zero.

The identification process did converge when four parameters were considered unknown ($\rho_{II}$, $G_{xz,III}$, $E_{x,IV}$ and $h_{IV}$) with the corresponding indices taken as in Table~\ref{tab:1}. The minimization process converged after 137 iterations each of which lasted approximately 14 s, resulting in a total computation time of 1950s on a laptop device. The properties identified through the results corresponding to the pressure wave are again presented in Table~\ref{tab:1}. Very good agreement exists between the recovered values and the ones initially injected in the full FE model (maximum divergence again not greater than 1\%), while the final value of the objective function was equal to 5.34. It is evident that incorrect wavenumber measurements will radically increase the value of the calculated objective function, therefore leading to a non-convergent problem. An experimental approach is employed in the following section in order to further extend the validation process.

\subsection{Structural identification through experimental measurements for a layered composite}\label{sec:3.2}

The experimental validation is based on experimental work by Droz et al.~\cite{ref30}. In this section, the proposed identification strategy is applied to a sandwich structure, and results are compared with the ones obtained in \cite{ref30} from IWC method, static experiments and the Transition Frequency Characterization.

\begin{figure}[htbp]
  \centering
  \includegraphics[width=0.50\linewidth]{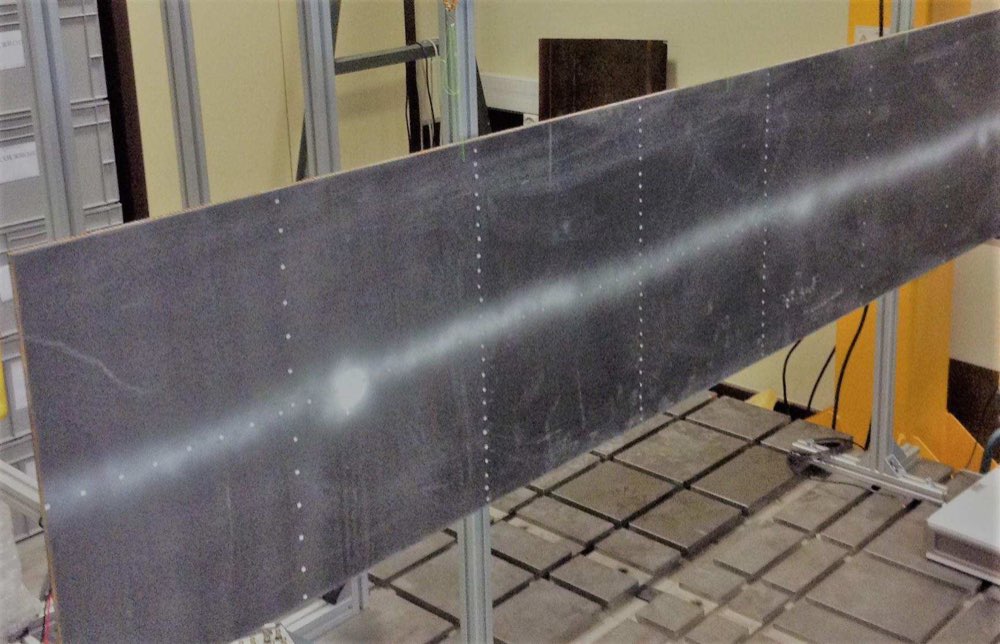}
  \caption{Photo of the structure used to retrieve experimental wavenumbers.}
  \label{fig:8}
\end{figure}

The structure is a rectangular sandwich plate measuring 60 cm $\times$ 288 cm, placed in a horizontal position as depicted in Fig.~\ref{fig:8}. The structural response is measured using a Polytec laser vibrometer and is transmitted to the acquisition system in order to compute the structural impedance at various points of the panel. The panel was excited using a shaker, controled by the Polytec acquisition system and adhered to the structure through a force sensor.

The constitutive materials are a 10 mm-thick Nomex honeycomb core involving a 3.2 mm cell size, while propagation is considered in the W-direction. The sandwich's skins are 0.6 mm-thick Hexforce with multi-axial, carbon-reinforced fibres. The density of the skins given by the manufacturer is $\rho_s = 1451$ kg m$^{-3}$ and the core's density is $\rho_c = 99$ kg m$^{-3}$.

Static measurements conducted on the layered structure provide the following mechanical characteristics for the Young's modulus of the facesheets and the shear modulus of the core along the investigated direction:
\begin{equation*}
E_{\mathrm{manuf}} = 70\ \mathrm{GPa} \quad \mathrm{and} \quad G_{\mathrm{manuf}} \in [30 - 38]\ \mathrm{MPa}
\tag{7}\label{eq:7}
\end{equation*}

\begin{figure}[htbp]
  \centering
  \includegraphics[width=0.75\linewidth]{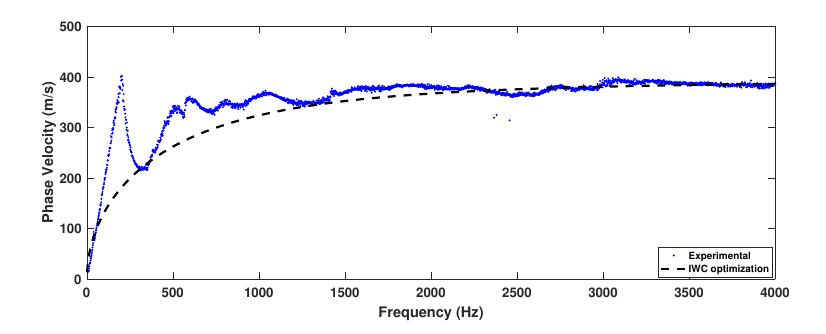}
  \caption{Flexural phase velocities obtained in the main direction of the plate. Inaccurate results are usually expected in the low frequencies for the IWC method. The convergence however increases with frequency, providing approximated material properties and a good correlation with analytical results.}
  \label{fig:9}
\end{figure}

The shaker is used to produce the harmonic excitation at the edge of the plate while the laser vibrometer is used to measure normal displacement field at the surface, along a 50 cm line of 66 points. The IWC method \cite{ref49} is employed to estimate the propagating flexural wavenumbers in the plate. The frequency range of interest spans 0 and 4000 Hz. The phase velocities obtained by the IWC method are shown in Fig.~\ref{fig:9}. Material properties obtained from the IWC characterization \cite{ref30} are:
\begin{equation*}
E_{\mathrm{IWC}} = 62\ \mathrm{GPa} \quad \mathrm{and} \quad G_{\mathrm{IWC}} = 37.8\ \mathrm{MPa}
\tag{8}\label{eq:8}
\end{equation*}

The method proposed in Section 2 is now applied to estimate the two mechanical characteristics mentioned above. Pulse measurements are post-processed to retrieve the phase velocities at selected frequencies. Note that the extraction of accurate phase velocities becomes easier in higher frequencies. Although wavenumbers can directly be retrieved

in the considered frequency range, this is not always the case in higher frequencies. Additionally, the 1 m distance required in this experimental set-up is due to the low wavenumbers of flexural waves in the sandwich panel. This distance could be considerably reduced if ultrasonic waves are employed to retrieve local dynamic properties.

\begin{figure}[htbp]
  \centering
  \includegraphics[width=0.80\linewidth]{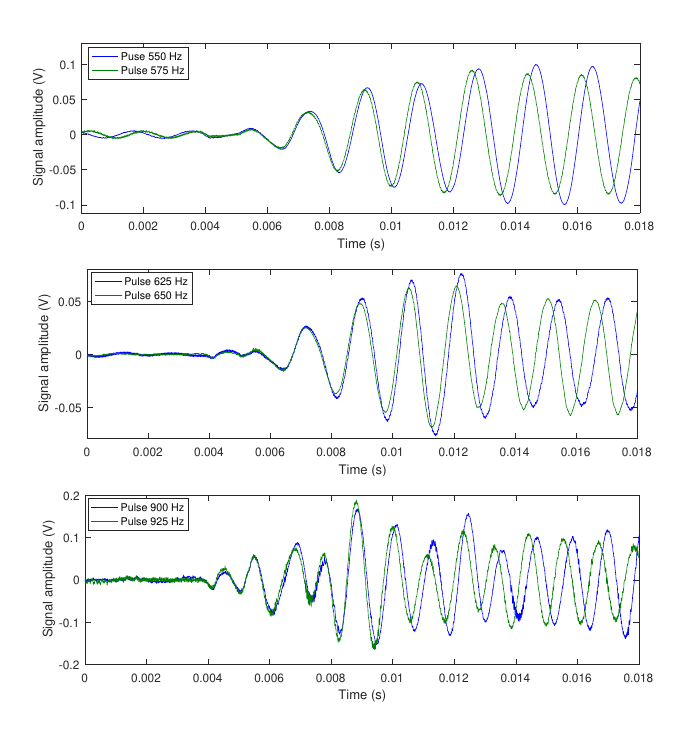}
  \caption{Measured pulse signals at acquisition point.}
  \label{fig:10}
\end{figure}

The pulse generation at frequency $f_n$ involves at least 10 cycles to limit dispersion effects occurring at these frequencies and is controlled by the PSV Laser using a triggering procedure. The response is measured at 1 m from the source for a selected number of frequencies between 500 Hz and 1500 Hz. The time signals are averaged at least 30 times to reduce experimental noise. Noteworthy, the group velocity can also be derived from measured phase velocities and wavenumbers. Measured pulses are shown in Fig.~\ref{fig:10} at 6 different frequencies close to the transition bandwidth.

Note that a refined frequency sampling was used in \cite{ref30} to evaluate the transition frequency at 880 Hz, and retrieve the following material characteristics:
\begin{equation*}
E_{\mathrm{TFC}} = 69.8\ \mathrm{GPa} \quad \mathrm{and} \quad G_{\mathrm{TFC}} = 36.5\ \mathrm{MPa}
\tag{9}\label{eq:9}
\end{equation*}

\begin{figure}[htbp]
  \centering
  \includegraphics[width=0.8\linewidth]{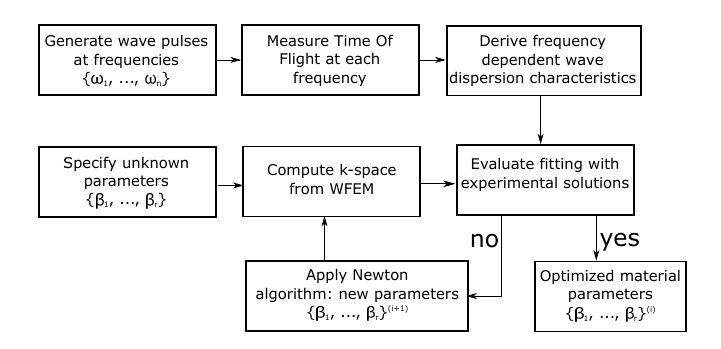}
  \caption{Experimental procedure for the WFE-based model updating strategy.}
  \label{fig:11}
\end{figure}

Taking into account the material characteristics provided by the manufacturer of the layered panel and presented above, the WFE iterative process is formed and the properties of the panel are identified through the presented, Newton-like minimization scheme. The process depicted in Fig.~\ref{fig:11} and detailed by Algorithm 1 was programmed and executed with the experimentally obtained flexural wavenumbers serving as the reference measured values. The structural parameters to be identified are the Young's modulus of the facesheet and the shear modulus of the core in the direction of wave propagation. A new design was therefore generated after each iteration, taking into account the first derivative of $\mathcal{F}$. After converging to a minimum of $\mathcal{F}$, the final value of the objective function was compared to the identification criterion. If the identification condition was not satisfied, a drastically altered design was evaluated by the iterative algorithm. Three elements comprise the WFE model which results in very fast model updating and eigenproblem solutions for Eq. (A.8). An identification criterion equal to 10 was employed while the minimization process converged in 91 iterations each of which lasted approximately 14 s, resulting in a total computation time of 1274s on a conventional laptop device.

The identified Young's modulus for the skins of the laminate and the shear modulus of the honeycomb core in the direction under investigation are computed as:
\begin{equation*}
E_{\mathrm{WFE}} = 69.5\ \mathrm{GPa} \quad \mathrm{and} \quad G_{\mathrm{WFE}} = 37.1\ \mathrm{MPa}
\tag{10}\label{eq:10}
\end{equation*}
which are both in very good agreement with the values provided by the other methods mentioned above, therefore experimentally validating the exhibited computational scheme.

\section{Conclusions}\label{sec:4}

In this work we have developed and applied a new identification technique based on FE modelling and the properties of propagating waves in multilayered structures. The principal contribution resulting from this work is a robust numerical NDE procedure for recovering effective structural parameters of complex, layered composites. It can be concluded that:

(i) The method is able to extract layer characteristics including thicknesses, densities, tensile and shear moduli for each individual layer and is robust enough to be applied in a broadband frequency range. Case studies elaborating on both ultrasonic as well as low frequency ranges were presented. In the ultrasound range the wave characteristics are straightforward to extract through the measured wave envelope, while in the low frequency regime dedicated techniques can be employed such as the IWC approach. (ii) The exhibited scheme was validated through comparison with experimental results as well as through full FE transient response predictions. Excellent agreement is observed for the identified structural parameters. (iii) It was shown that faster convergence of the post-processing identification algorithm was attained within the so-called wavenumber transition spectrum where the bending-to-shear transition phenomenon can be easily captured using the WFE method. (iv) It is emphasized that the proposed wave-based method has significant advantages compared to modal identification approaches. More precisely the accuracy of the structural parameters is not altered by the presence of uncertain boundaries since the data is obtained locally, through single-shot measurements. This is a considerable advantage compared to a number of stationary and other existing methods, since it can then be applied in situ and without requiring additional sampling of the structure. The use of unlimited and user-selected excitation frequencies can effectively increase the number of parameters to be identified through the inverse wave modelling, resulting in a significant increase of the method's robustness in a broadband frequency sense. (v) The principal drawback of the presented approach is the required computational effort. This can range from negligible (when a single structural parameter is sought) to intense, when typically more than three parameters are to be identified and several iterations need to be completed before the Newton's scheme converges to the desired solution. Providing expressions of the wavenumber sensitivity to the investigated structural parameters under investigation can accelerate the convergence process. Drastic computational savings can be attained by a-priori solving the WFE forward model for a fine grid of variables and using a neural-network type approach for extracting the desired parameters; this is currently a topic of further research.

\section*{Acknowledgements}\label{sec:ack}

The authors would like to gratefully acknowledge the H2020 ITN Marie-Curie project GA-721455 'SAFE-FLY EID', as well as the EPSRC thematic doctoral programme 'Waves in Complex Media' for the financial support.

\section*{Appendix A. The forward wave and finite element modelling approach}\label{app:A}

Linear elastic wave propagation is considered in the $x$ direction of the arbitrarily layered structural waveguide of Fig.~\ref{fig:1}. The problem can be condensed using a transfer matrix approach as in \cite{ref33}. The frequency dependent Dynamic Stiffness Matrix (DMS) of the waveguide's periodic segment can be partitioned with regard to its left/right sides and internal DoF as
\begin{equation*}
\begin{bmatrix} \mathbf{D}_{LL} & \mathbf{D}_{LI} & \mathbf{D}_{LR} \\ \mathbf{D}_{IL} & \mathbf{D}_{II} & \mathbf{D}_{IR} \\ \mathbf{D}_{RL} & \mathbf{D}_{RI} & \mathbf{D}_{RR} \end{bmatrix} \begin{Bmatrix} \mathbf{q}_{L} \\ \mathbf{q}_{I} \\ \mathbf{q}_{R} \end{Bmatrix} = \begin{Bmatrix} \mathbf{f}_{L} \\ \mathbf{0} \\ \mathbf{f}_{R} \end{Bmatrix}
\tag{A.1}\label{eq:A.1}
\end{equation*}
with $\mathbf{q}$ the displacement and $\mathbf{f}$ the forcing vectors. Using a Guyan-type condensation for the internal DoF the problem can be expressed as
\begin{equation*}
\begin{bmatrix} \mathbf{D}_{LL} - \mathbf{D}_{LI} \mathbf{D}_{II}^{-1} \mathbf{D}_{IL} & \mathbf{D}_{LR} - \mathbf{D}_{LI} \mathbf{D}_{II}^{-1} \mathbf{D}_{IR} \\ \mathbf{D}_{RL} - \mathbf{D}_{RI} \mathbf{D}_{II}^{-1} \mathbf{D}_{IL} & \mathbf{D}_{RR} - \mathbf{D}_{RI} \mathbf{D}_{II}^{-1} \mathbf{D}_{IR} \end{bmatrix} \begin{Bmatrix} \mathbf{q}_{L} \\ \mathbf{q}_{R} \end{Bmatrix} = \begin{Bmatrix} \mathbf{f}_{L} \\ \mathbf{f}_{R} \end{Bmatrix}
\tag{A.2}\label{eq:A.2}
\end{equation*}

Assuming that no external forces are applied on the segment the displacement continuity and force equilibrium equations at the interface of two consecutive periodic segments $s$ and $s+1$ give
\begin{equation*}
\begin{aligned} \mathbf{q}_L^{s+1} &= \mathbf{q}_R^s \\ \mathbf{f}_L^{s+1} &= -\mathbf{f}_R^s \end{aligned}
\tag{A.3}\label{eq:A.3}
\end{equation*}

Using Eqs. (A.2) and (A.3) the relation of the displacements and forces of the left and right sides of the segment can be written as
\begin{equation*}
\begin{Bmatrix} \mathbf{q}_{L}^{s+1} \\ \mathbf{f}_{L}^{s+1} \end{Bmatrix} = \mathbf{T} \begin{Bmatrix} \mathbf{q}_{L}^{s} \\ \mathbf{f}_{L}^{s} \end{Bmatrix}
\tag{A.4}\label{eq:A.4}
\end{equation*}
and the expression of the symplectic transfer matrix $\mathbf{T}$ can be written as
\begin{equation*}
\mathbf{T} = \begin{bmatrix} \mathbf{D}_{11} & \mathbf{D}_{12} \\ \mathbf{D}_{21} & \mathbf{D}_{22} \end{bmatrix}_{[2j \times 2j]}
\tag{A.5}\label{eq:A.5}
\end{equation*}
with
\begin{equation*}
\begin{aligned}
\mathbf{D}_{11} &= -(\mathbf{D}_{LR} - \mathbf{D}_{LI} \mathbf{D}_{II}^{-1} \mathbf{D}_{IR})^{-1} (\mathbf{D}_{LL} - \mathbf{D}_{LI} \mathbf{D}_{II}^{-1} \mathbf{D}_{IL}) \\
\mathbf{D}_{12} &= (\mathbf{D}_{LR} - \mathbf{D}_{LI} \mathbf{D}_{II}^{-1} \mathbf{D}_{IR})^{-1} \\
\mathbf{D}_{21} &= -\mathbf{D}_{RL} + \mathbf{D}_{RI} \mathbf{D}_{II}^{-1} \mathbf{D}_{IL} + (\mathbf{D}_{RR} - \mathbf{D}_{RI} \mathbf{D}_{II}^{-1} \mathbf{D}_{IR}) (\mathbf{D}_{LR} - \mathbf{D}_{LI} \mathbf{D}_{II}^{-1} \mathbf{D}_{IR})^{-1} (\mathbf{D}_{LL} - \mathbf{D}_{LI} \mathbf{D}_{II}^{-1} \mathbf{D}_{IL}) \\
\mathbf{D}_{22} &= -(\mathbf{D}_{RR} - \mathbf{D}_{RI} \mathbf{D}_{II}^{-1} \mathbf{D}_{IR}) (\mathbf{D}_{LR} - \mathbf{D}_{LI} \mathbf{D}_{II}^{-1} \mathbf{D}_{IR})^{-1}
\end{aligned}
\tag{A.6}\label{eq:A.6}
\end{equation*}

With a wave propagating freely along the $x$ direction, the propagation constant $\gamma = \mathrm{e}^{-\mathrm{i}kL_x}$ relates the right and left nodal displacements and forces by
\begin{equation*}
\begin{aligned} \mathbf{q}_R^s &= \gamma \mathbf{q}_L^s \\ \mathbf{f}_R^s &= -\gamma \mathbf{f}_L^s \end{aligned}
\tag{A.7}\label{eq:A.7}
\end{equation*}

By substituting Eqs. (A.3) and (A.7) in Eq. (A.4), the free wave propagation is described by the eigenproblem
\begin{equation*}
\gamma \begin{Bmatrix} \mathbf{q}_{L}^{s} \\ \mathbf{f}_{L}^{s} \end{Bmatrix} = \mathbf{T} \begin{Bmatrix} \mathbf{q}_{L}^{s} \\ \mathbf{f}_{L}^{s} \end{Bmatrix}
\tag{A.8}\label{eq:A.8}
\end{equation*}
whose eigenvalues $\gamma_w$ and eigenvectors $\boldsymbol{\phi}_w = \begin{Bmatrix} \boldsymbol{\phi}_q \\ \boldsymbol{\phi}_f \end{Bmatrix}_w$ solution sets provide a comprehensive description of the propagation constants and the wave mode shapes for each of the elastic waves propagating in the structural waveguide at a specified angular frequency $\omega$. Both positive going (with $\gamma_w^+$ and $\boldsymbol{\phi}_w^+$) and negative going waves ($\gamma_w^-$ and $\boldsymbol{\phi}_w^-$) are sought through the eigensolution. Positive going waves are characterised \cite{ref50} by
\begin{equation*}
\begin{gathered} |\gamma_w^+| \leq 1, \\ \Re(\mathrm{i}\omega \boldsymbol{\phi}_f^{+\top} \boldsymbol{\phi}_q^+) < 0 \text{ if } |\gamma_w^+| = 1 \end{gathered}
\tag{A.9}\label{eq:A.9}
\end{equation*}
stating that when a wave is travelling in the positive $x$ direction its amplitude should be decreasing, or that if its amplitude remains constant (in the case of propagating waves with complete absence of attenuation), then there is time averaged power transmission in the positive direction.

\section*{Appendix B. Structural FE matrices expressed directly as a function of layer mechanical characteristics}\label{app:B}

\begin{figure}[htbp]
  \centering
  \includegraphics[width=0.40\linewidth]{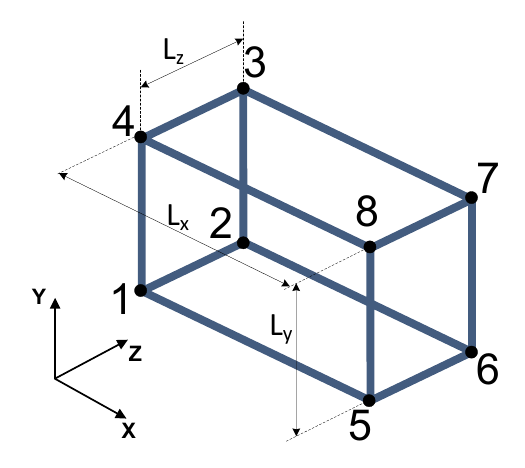}
  \caption{The considered cuboid solid FE}
  \label{fig:12}
\end{figure}

A linear solid FE is hereby considered as shown in Fig.~\ref{fig:12}. Following the isoparametric notation introduced in \cite{ref51} the geometry of the element is described as
\begin{equation*}
\begin{Bmatrix} x \\ y \\ z \end{Bmatrix} = \begin{bmatrix} x_{1} & x_{2} & x_{3} & x_{4} & x_{5} & x_{6} & x_{7} & x_{8} \\ y_{1} & y_{2} & y_{3} & y_{4} & y_{5} & y_{6} & y_{7} & y_{8} \\ z_{1} & z_{2} & z_{3} & z_{4} & z_{5} & z_{6} & z_{7} & z_{8} \end{bmatrix} \begin{Bmatrix} N_{1} \\ N_{2} \\ N_{3} \\ N_{4} \\ N_{5} \\ N_{6} \\ N_{7} \\ N_{8} \end{Bmatrix}
\tag{B.1}\label{eq:B.1}
\end{equation*}

The displacement interpolations are expressed as

\begin{equation*}
\begin{Bmatrix} u_{x} \\ u_{y} \\ u_{z} \end{Bmatrix} = \begin{bmatrix} u_{x1} & u_{x2} & u_{x3} & u_{x4} & u_{x5} & u_{x6} & u_{x7} & u_{x8} \\ u_{y1} & u_{y2} & u_{y3} & u_{y4} & u_{y5} & u_{y6} & u_{y7} & u_{y8} \\ u_{z1} & u_{z2} & u_{z3} & u_{z4} & u_{z5} & u_{z6} & u_{z7} & u_{z8} \end{bmatrix} \begin{Bmatrix} N_{1} \\ N_{2} \\ N_{3} \\ N_{4} \\ N_{5} \\ N_{6} \\ N_{7} \\ N_{8} \end{Bmatrix}
\tag{B.2}\label{eq:B.2}
\end{equation*}

Linear shape functions are assumed for the element
\begin{equation*}
\begin{aligned}
N_{1} &= \tfrac{1}{8}(1 - \xi)(1 - \eta)(1 + \mu) \\
N_{2} &= \tfrac{1}{8}(1 - \xi)(1 - \eta)(1 - \mu) \\
N_{3} &= \tfrac{1}{8}(1 - \xi)(1 + \eta)(1 - \mu) \\
N_{4} &= \tfrac{1}{8}(1 - \xi)(1 + \eta)(1 + \mu) \\
N_{5} &= \tfrac{1}{8}(1 + \xi)(1 - \eta)(1 + \mu) \\
N_{6} &= \tfrac{1}{8}(1 + \xi)(1 - \eta)(1 - \mu) \\
N_{7} &= \tfrac{1}{8}(1 + \xi)(1 + \eta)(1 - \mu) \\
N_{8} &= \tfrac{1}{8}(1 + \xi)(1 + \eta)(1 + \mu)
\end{aligned}
\tag{B.3}\label{eq:B.3}
\end{equation*}

The element stiffness matrix $\mathbf{k}$ is formally given by the volume integral
\begin{equation*}
\mathbf{k} = \int_{-1}^{1} \int_{-1}^{1} \int_{-1}^{1} \mathbf{B}^{\top} \mathbf{D} \mathbf{B} |\mathbf{J}| \, \mathrm{d}\eta \mathrm{d}\xi \mathrm{d}\mu
\tag{B.4}\label{eq:B.4}
\end{equation*}
while the element mass and damping matrices $\mathbf{m}$, $\mathbf{c}$ can be determined as
\begin{equation*}
\mathbf{m} = \int_{-1}^{1} \int_{-1}^{1} \int_{-1}^{1} \mathbf{N}^{\top} \rho_{m} \mathbf{N} |\mathbf{J}| \, \mathrm{d}\eta \mathrm{d}\xi \mathrm{d}\mu
\tag{B.5}\label{eq:B.5}
\end{equation*}

\begin{equation*}
\mathbf{c} = \int_{-1}^{1} \int_{-1}^{1} \int_{-1}^{1} \mathbf{N}^{\top} \gamma \, \mathbf{N} |\mathbf{J}| \, \mathrm{d}\eta \mathrm{d}\xi \mathrm{d}\mu
\tag{B.6}\label{eq:B.6}
\end{equation*}
with
\begin{equation*}
\mathbf{N} = \begin{bmatrix} N_1 & 0 & 0 & \cdots & N_8 & 0 & 0 \\ 0 & N_1 & 0 & \cdots & 0 & N_8 & 0 \\ 0 & 0 & N_1 & \cdots & 0 & 0 & N_8 \end{bmatrix}
\tag{B.7}\label{eq:B.7}
\end{equation*}
while $\rho_m$ is the mass density of the material and $\gamma$ the material damping coefficient. It is also noted that
\begin{equation*}
\mathbf{B} = \begin{bmatrix} \frac{\partial N_{1}}{\partial x} & 0 & 0 & \frac{\partial N_{2}}{\partial x} & \cdots & \frac{\partial N_{8}}{\partial x} & 0 & 0 \\ 0 & \frac{\partial N_{1}}{\partial y} & 0 & 0 & \cdots & 0 & \frac{\partial N_{8}}{\partial y} & 0 \\ 0 & 0 & \frac{\partial N_{1}}{\partial z} & 0 & \cdots & 0 & 0 & \frac{\partial N_{8}}{\partial z} \\ \frac{\partial N_{1}}{\partial y} & \frac{\partial N_{1}}{\partial x} & 0 & \frac{\partial N_{2}}{\partial y} & \cdots & \frac{\partial N_{8}}{\partial y} & \frac{\partial N_{8}}{\partial x} & 0 \\ 0 & \frac{\partial N_{1}}{\partial z} & \frac{\partial N_{1}}{\partial y} & 0 & \cdots & 0 & \frac{\partial N_{8}}{\partial z} & \frac{\partial N_{8}}{\partial y} \\ \frac{\partial N_{1}}{\partial z} & 0 & \frac{\partial N_{1}}{\partial x} & \frac{\partial N_{2}}{\partial z} & \cdots & \frac{\partial N_{8}}{\partial z} & 0 & \frac{\partial N_{8}}{\partial x} \end{bmatrix} \begin{Bmatrix} u_{x1} \\ u_{y1} \\ u_{z1} \\ u_{x2} \\ \cdots \\ u_{x8} \\ u_{y8} \\ u_{z8} \end{Bmatrix}
\tag{B.8}\label{eq:B.8}
\end{equation*}

The Jacobian matrix of the element is
\begin{equation*}
\mathbf{J} = \begin{bmatrix} \frac{\partial x}{\partial \xi} & \frac{\partial y}{\partial \xi} & \frac{\partial z}{\partial \xi} \\ \frac{\partial x}{\partial \eta} & \frac{\partial y}{\partial \eta} & \frac{\partial z}{\partial \eta} \\ \frac{\partial x}{\partial \mu} & \frac{\partial y}{\partial \mu} & \frac{\partial z}{\partial \mu} \end{bmatrix}
\tag{B.9}\label{eq:B.9}
\end{equation*}
while the flexibility matrix of the element for an orthotropic material $\mathbf{D}^{-1}$ can generally be written as
\begin{equation*}
\mathbf{D}^{-1} = \begin{bmatrix} \frac{1}{E_x} & \frac{-v_{xy}}{E_x} & \frac{-v_{xz}}{E_x} & 0 & 0 & 0 \\ \frac{-v_{yx}}{E_y} & \frac{1}{E_y} & \frac{-v_{yz}}{E_y} & 0 & 0 & 0 \\ \frac{-v_{zx}}{E_z} & \frac{-v_{zy}}{E_z} & \frac{1}{E_z} & 0 & 0 & 0 \\ 0 & 0 & 0 & \frac{1}{G_{xy}} & 0 & 0 \\ 0 & 0 & 0 & 0 & \frac{1}{G_{yz}} & 0 \\ 0 & 0 & 0 & 0 & 0 & \frac{1}{G_{xz}} \end{bmatrix}
\tag{B.10}\label{eq:B.10}
\end{equation*}

The assumption of the undeformed FE being a rectangular parallelepiped is hereby adopted. The coordinates $x_1, x_2, x_3, x_4, x_5, x_6, x_7, x_8$, $y_1, y_2, y_3, y_4, y_5, y_6, y_7, y_8$, and $z_1, z_2, z_3, z_4, z_5, z_6, z_7, z_8$, can then be replaced by $L_x, L_y, L_z$ in the expression of $\mathbf{B}$. The generic expression for $\mathbf{m}$ is thus given as
\begin{multline*}
\mathbf{m} = (\rho L_{x} L_{y} L_{z}) \\
\left[ \text{\tiny $\setlength{\arraycolsep}{0.5pt}\begin{array}{*{24}{c}}
1/27 & 0 & 0 & 1/54 & 0 & 0 & 1/108 & 0 & 0 & 1/54 & 0 & 0 & 1/54 & 0 & 0 & 1/108 & 0 & 0 & 1/216 & 0 & 0 & 1/108 & 0 & 0 \\
0 & 1/27 & 0 & 0 & 1/54 & 0 & 0 & 1/108 & 0 & 0 & 1/54 & 0 & 0 & 1/54 & 0 & 0 & 1/108 & 0 & 0 & 1/216 & 0 & 0 & 1/108 & 0 \\
0 & 0 & 1/27 & 0 & 0 & 1/54 & 0 & 0 & 1/108 & 0 & 0 & 1/54 & 0 & 0 & 1/54 & 0 & 0 & 1/108 & 0 & 0 & 1/216 & 0 & 0 & 1/108 \\
1/54 & 0 & 0 & 1/27 & 0 & 0 & 1/54 & 0 & 0 & 1/108 & 0 & 0 & 1/108 & 0 & 0 & 1/54 & 0 & 0 & 1/108 & 0 & 0 & 1/216 & 0 & 0 \\
0 & 1/54 & 0 & 0 & 1/27 & 0 & 0 & 1/54 & 0 & 0 & 1/108 & 0 & 0 & 1/108 & 0 & 0 & 1/54 & 0 & 0 & 1/108 & 0 & 0 & 1/216 & 0 \\
0 & 0 & 1/54 & 0 & 0 & 1/27 & 0 & 0 & 1/54 & 0 & 0 & 1/108 & 0 & 0 & 1/108 & 0 & 0 & 1/54 & 0 & 0 & 1/108 & 0 & 0 & 1/216 \\
1/108 & 0 & 0 & 1/54 & 0 & 0 & 1/27 & 0 & 0 & 1/54 & 0 & 0 & 1/216 & 0 & 0 & 1/108 & 0 & 0 & 1/54 & 0 & 0 & 1/108 & 0 & 0 \\
0 & 1/108 & 0 & 0 & 1/54 & 0 & 0 & 1/27 & 0 & 0 & 1/54 & 0 & 0 & 1/216 & 0 & 0 & 1/108 & 0 & 0 & 1/54 & 0 & 0 & 1/108 & 0 \\
0 & 0 & 1/108 & 0 & 0 & 1/54 & 0 & 0 & 1/27 & 0 & 0 & 1/54 & 0 & 0 & 1/216 & 0 & 0 & 1/108 & 0 & 0 & 1/54 & 0 & 0 & 1/108 \\
1/54 & 0 & 0 & 1/108 & 0 & 0 & 1/54 & 0 & 0 & 1/27 & 0 & 0 & 1/108 & 0 & 0 & 1/216 & 0 & 0 & 1/108 & 0 & 0 & 1/54 & 0 & 0 \\
0 & 1/54 & 0 & 0 & 1/108 & 0 & 0 & 1/54 & 0 & 0 & 1/27 & 0 & 0 & 1/108 & 0 & 0 & 1/216 & 0 & 0 & 1/108 & 0 & 0 & 1/54 & 0 \\
0 & 0 & 1/54 & 0 & 0 & 1/108 & 0 & 0 & 1/54 & 0 & 0 & 1/27 & 0 & 0 & 1/108 & 0 & 0 & 1/216 & 0 & 0 & 1/108 & 0 & 0 & 1/54 \\
1/54 & 0 & 0 & 1/108 & 0 & 0 & 1/216 & 0 & 0 & 1/108 & 0 & 0 & 1/27 & 0 & 0 & 1/54 & 0 & 0 & 1/108 & 0 & 0 & 1/54 & 0 & 0 \\
0 & 1/54 & 0 & 0 & 1/108 & 0 & 0 & 1/216 & 0 & 0 & 1/108 & 0 & 0 & 1/27 & 0 & 0 & 1/54 & 0 & 0 & 1/108 & 0 & 0 & 1/54 & 0 \\
0 & 0 & 1/54 & 0 & 0 & 1/108 & 0 & 0 & 1/216 & 0 & 0 & 1/108 & 0 & 0 & 1/27 & 0 & 0 & 1/54 & 0 & 0 & 1/108 & 0 & 0 & 1/54 \\
1/108 & 0 & 0 & 1/54 & 0 & 0 & 1/108 & 0 & 0 & 1/216 & 0 & 0 & 1/54 & 0 & 0 & 1/27 & 0 & 0 & 1/54 & 0 & 0 & 1/108 & 0 & 0 \\
0 & 1/108 & 0 & 0 & 1/54 & 0 & 0 & 1/108 & 0 & 0 & 1/216 & 0 & 0 & 1/54 & 0 & 0 & 1/27 & 0 & 0 & 1/54 & 0 & 0 & 1/108 & 0 \\
0 & 0 & 1/108 & 0 & 0 & 1/54 & 0 & 0 & 1/108 & 0 & 0 & 1/216 & 0 & 0 & 1/54 & 0 & 0 & 1/27 & 0 & 0 & 1/54 & 0 & 0 & 1/108 \\
1/216 & 0 & 0 & 1/108 & 0 & 0 & 1/54 & 0 & 0 & 1/108 & 0 & 0 & 1/108 & 0 & 0 & 1/54 & 0 & 0 & 1/27 & 0 & 0 & 1/54 & 0 & 0 \\
0 & 1/216 & 0 & 0 & 1/108 & 0 & 0 & 1/54 & 0 & 0 & 1/108 & 0 & 0 & 1/108 & 0 & 0 & 1/54 & 0 & 0 & 1/27 & 0 & 0 & 1/54 & 0 \\
0 & 0 & 1/216 & 0 & 0 & 1/108 & 0 & 0 & 1/54 & 0 & 0 & 1/108 & 0 & 0 & 1/108 & 0 & 0 & 1/54 & 0 & 0 & 1/27 & 0 & 0 & 1/54 \\
1/108 & 0 & 0 & 1/216 & 0 & 0 & 1/108 & 0 & 0 & 1/54 & 0 & 0 & 1/54 & 0 & 0 & 1/108 & 0 & 0 & 1/54 & 0 & 0 & 1/27 & 0 & 0 \\
0 & 1/108 & 0 & 0 & 1/216 & 0 & 0 & 1/108 & 0 & 0 & 1/54 & 0 & 0 & 1/54 & 0 & 0 & 1/108 & 0 & 0 & 1/54 & 0 & 0 & 1/27 & 0 \\
0 & 0 & 1/108 & 0 & 0 & 1/216 & 0 & 0 & 1/108 & 0 & 0 & 1/54 & 0 & 0 & 1/54 & 0 & 0 & 1/108 & 0 & 0 & 1/54 & 0 & 0 & 1/27
\end{array}$} \right] \\
\tag{B.11}\label{eq:B.11}
\end{multline*}
a very similar expression is true for $\mathbf{c}$, while the symbolic generic expression of $\mathbf{k}$ can be derived exactly in the same way but is hereby intentionally omitted for the sake of brevity. This implies the very practical fact of the mass, damping and stiffness matrices $\mathbf{m}$ and $\mathbf{k}$ for each independent layer of the structure being a direct expression of the structural and material characteristics $\rho$, $L_z$, $E_x$, $E_y$, $E_z$, $G_{xy}$, $G_{xz}$, $G_{zy}$, $v_{xy}$, $v_{xz}$, $v_{zy}$. It is obvious that for an isotropic layer, the above expressions are radically simplified, while the stiffness and mass matrices of a multilayer structure can be obtained by assembling the discrete layer matrices. A Newton-like iterative scheme can then be effectively employed for minimising the objective function and solving the system of eigenvalue expressions for recovering the mechanical characteristics of each individual layer.

The generic sensitivity expressions $\frac{\partial \mathbf{k}}{\partial \beta_i}$, $\frac{\partial \mathbf{m}}{\partial \beta_i}$ as well as $\frac{\partial^2 \mathbf{k}}{\partial \beta_j \partial \beta_i}$, $\frac{\partial^2 \mathbf{m}}{\partial \beta_j \partial \beta_i}$ with $\beta_i$, $\beta_j$ being design parameters can therefore be calculated as a function of $E_x$, $E_y$, $E_z$, $v_{xy}$, $v_{xz}$, $v_{yz}$, $G_{xy}$, $G_{xz}$, $G_{yz}$, $L_x$, $L_y$, $L_z$ by differentiating over the generic expressions for $\mathbf{k}$, $\mathbf{m}$. The sensitivities of the global matrices $\frac{\partial \mathbf{K}}{\partial \beta_i}$, $\frac{\partial \mathbf{M}}{\partial \beta_i}$ can then be computed by assembling the individual element sensitivity matrices together.

\end{document}